\documentclass[twocolumn,showpacs,preprintnumbers,amsmath,amssymb,superscriptaddress]{revtex4}

\usepackage{graphicx}
\usepackage{axodraw}
\usepackage{epsfig}
\usepackage{bm}
\def\mathswitchr#1{\relax\ifmmode{\mathrm{#1}}\else$\mathrm{#1}$\fi}
\newcommand{\Pp}{\mathswitchr {p}}
\newcommand{\Pg}{\mathswitchr {g}}
\newcommand{\Pf}{\mathswitchr {f}}
\newcommand{\Pq}{\mathswitchr {q}}
\newcommand{\Pu}{\mathswitchr {u}}
\newcommand{\Pd}{\mathswitchr {d}}
\newcommand{\Pc}{\mathswitchr {c}}
\newcommand{\Ps}{\mathswitchr {s}}
\newcommand{\Pt}{\mathswitchr {t}}
\newcommand{\Pb}{\mathswitchr {b}}
\newcommand{\Pe}{\mathswitchr {e}}
\newcommand{\Pl}{\mathswitchr {l}}
\newcommand{\Pcm}{\mathswitchr {cm}}
\newcommand{\PV}{\mathswitchr {V}}
\newcommand{\PW}{\mathswitchr {W}}
\newcommand{\PZ}{\mathswitchr {Z}}
\newcommand{\Pleptons}{\mathswitchr {leptons}}
\newcommand{\cm}{\mathswitchr {cm}}
\newcommand{\cut}{\mathswitchr {cut}}
\newcommand{\PBox}{\mathswitchr {Box}}
\newcommand{\Pcos}{\mathswitchr {cos}}
\newcommand{\SM}{\mathswitchr {SM}}
\newcommand{\noH}{\mathswitchr {noH}}
\newcommand{\GeV}{\unskip\,\mathrm{GeV}}

\newcommand{\TeV}{\unskip\,\mathrm{TeV}}
\newcommand{\fba}{\unskip\,\mathrm{fb}}

\def\mathswitch#1{\relax\ifmmode#1\else$#1$\fi}
\newcommand{\MW}{\mathswitch {M_\PW}}
\newcommand{\MZ}{\mathswitch {M_\PZ}}
\newcommand{\Mb}{\mathswitch {m_\Pb}}
\newcommand{\Mt}{\mathswitch {m_\Pt}}
\newcommand{\Mll}{\mathswitch {M_{\Pl\Pl^\prime}}}
\newcommand{\GW}{\mathswitch {\Gamma_\PW}}
\newcommand{\GZ}{\Gamma_{\PZ}}
\newcommand{\PT}{P_{\mathrm{T}}}
\newcommand{\PTmiss}{P_{\mathrm{T}}^{\mathrm{miss}}}

\def\si{\sigma}
\newcommand{\scrs}{\scriptscriptstyle}
\newcommand{\sw}{\mathswitch {s_{\scrs\PW}}}
\def\beq{\begin{equation}}
\def\eeq{\end{equation}}
\def\beqar{\begin{eqnarray}}
\def\eeqar{\end{eqnarray}}
\def\nl{\nonumber\\}
\def\ie{i.e.\ }
\newcommand{\rd}{{\mathrm{d}}}

\begin{document}
\title{The process {\boldmath $gg\to WW$} as a probe into the EWSB mechanism}
\author{E.~Accomando}
\affiliation{INFN, Torino \& Dipartimento di Fisica Teorica, 
                   Universit\`a di Torino, I-10125 Torino, Italy}
\date{September 3, 2007}
\begin{abstract}
We present the first estimate of the gluon-gluon fusion process, 
$\Pg\Pg\to 4\Pf$, in the high energy domain as a probe of the electroweak 
symmetry breaking mechanism (EWSB). We consider the exact matrix element at 
${\cal O}(\alpha_s\alpha_{em}^2)$, and we include all irreducible background 
coming from $\Pq\bar\Pq\rightarrow 4\Pf$. Purely leptonic final states, 
$\Pp\Pp\to l\bar\nu_l\nu_{l^\prime}\bar{l^\prime}$, are numerically 
investigated. We find that this channel is extremely sensitive to the regime 
of the interaction between gauge bosons. It can thus be associated to the 
traditionally used vector boson scattering (VBS) to improve the analysis of 
the EWSB physics.
\end{abstract}

                             
\maketitle

The discovery of the EWSB physics will be the primary goal of the LHC. This 
Letter deals with the study of a new process, which could largely improve the 
LHC potential in this search. We consider the production of $\PW\PW$-pairs via 
gluon-gluon fusion, $gg\to\PW\PW\to 4\Pf$. Usually analysed for the Higgs 
boson discovery, \ie in the low-intermediate energy range where the Higgs 
resonance is expected to appear, this channel is here found to have a strong 
potential also at high energies. Our aim is to present the properties of the 
gluon-induced weak-boson pair production at the $\TeV$ scale, and to analyse 
their consequences on the phenomenology of the interaction between the 
produced gauge-bosons. The main motivation for such a study relies on the 
strict correlation between the regime of the gauge interaction and the 
mechanism which triggers the EWSB\cite{ewsb}.

Many theories describe different EWSB scenarios. Most of them (Standard Model 
(SM), SUSY, etc.) predict the existence of at least one light Higgs. This 
hypotesis implies that the dynamics responsible of the spontaneous symmetry 
breaking is weakly-coupled. Such a picture is in good agreement with the LEP1 
electroweak precision measurements. 
However, the Higgs is still missing. In addition, new theoretical developments 
have opened up the possibility to build new models of electroweak symmetry 
breaking (a recent review is given in Ref.~\cite{rattazzi:2006}). They mainly 
fall into two classes. In the first case, the Higgs is
still predicted but it is not an elementary particle. It is included as an
effective field arising from a new dynamics which becomes strong at some
energy scale (an example are the Little Higgs models).   
In the latter, the Higgs sector might even be completely replaced with 
strongly interacting dynamics. Interesting realizations of this scenario can 
arise in extra-dimensions theories.

Hence, in order to understand the nature of the new physics which will be 
discovered in the next future, a crucial issue to be settled is whether the 
EWSB dynamics is weakly or strongly coupled. 
A way of answering this question preserving a model independent approach is 
thus looking at processes involving at least one massive gauge-boson pair.
Ideally, the vector boson scattering, $\PV\PV\to \PV\PV$ ($\PV=\PW,\PZ$), is 
the most sensitive process to the EWSB mechanism \cite{ewsb}. However, it is 
embedded in 
the more complex channel $\Pq_1\Pq_2\to\Pq_3\Pq_4\PV\PV\to\Pq_3\Pq_4+4\Pf$.
Its sensitivity is thus depleted by limited number of events and huge 
backgrounds. For a recent and detailed status of VBS perspectives at the LHC 
see for instance Ref.~\cite{vbs} and references therein.  
The gluon-induced process can bring a powerful help in this challenge.

\begin{figure}[t]
\begin{center}
\includegraphics[width=4.2cm,angle=0]{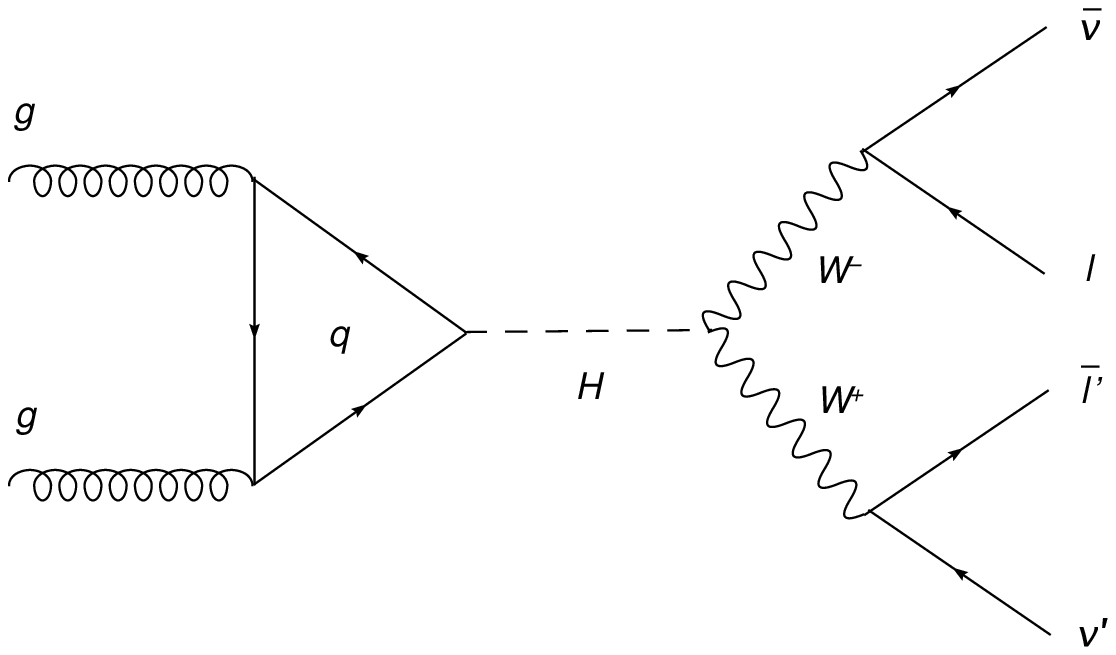}~~~~~
\includegraphics[width=3.5cm,angle=0]{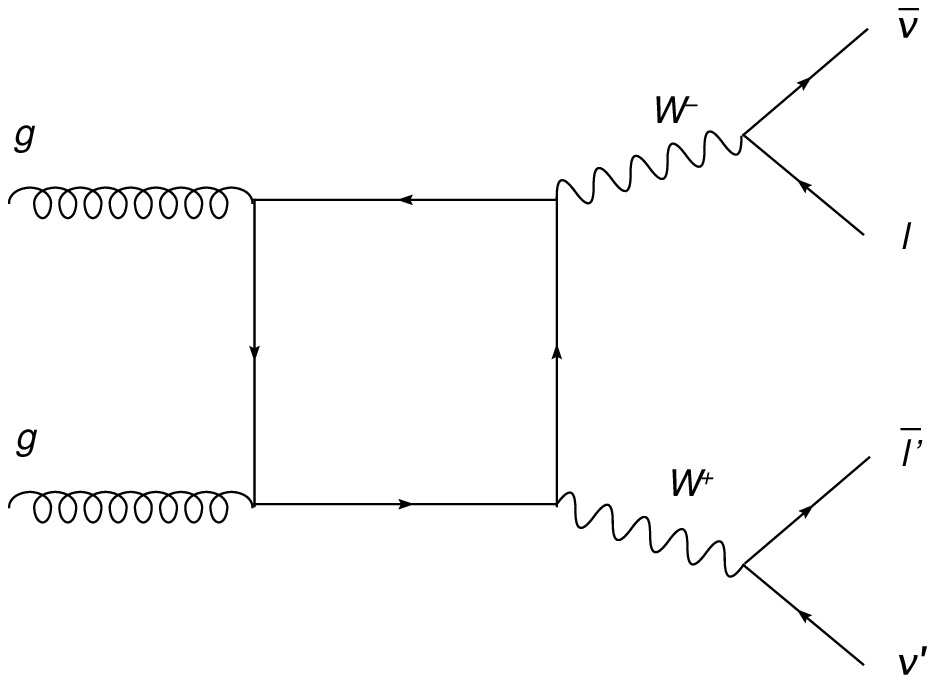}
\vspace{-.2cm}
\caption{Sample LO diagram for the $\Pg\Pg\to 4\Pf$ process.
\vspace{-.7cm}
\label{fig:LOdiag}} 
\end{center} 
\end{figure}
Let us begin with recalling that the leading-order (LO) contribution to the 
gluon-induced weak-boson pair production originates from one-loop diagrams in 
the SM. An example is shown in Fig.~\ref{fig:LOdiag} \footnote {For massless 
external particles, diagrams in Fig.~\ref{fig:LOdiag} are the only 
ones to contribute owing to the Furry's theorem.}. The corresponding 
amplitudes were first evaluated
in the on-shell vector-boson approximation, $\Pg\Pg\to\PV\PV$ 
\cite{Glover:1989,Kao:1991}. Successive computations took into account the
spin correlations between gauge-boson production and decay, by considering 
the process $\Pg\Pg\to\PV\PV\to 4\Pf$ in narrow width approximation 
\cite{Duhrssen:2005-Matsuura:1991-Zecher:1994}. 
Recently, the first full calculation of the loop-induced $\PW$-boson pair 
production and decay, $\Pg\Pg\to 4\Pleptons$, was implemented in the code 
$\tt gg2WW$ and published \cite{Binoth:2006-Binoth:2005}.
It takes into account vector-boson off-shell effects, full spin and decay 
angle correlations, and the loop contribution coming from the massive third 
generation quarks. 

Starting from this result, we have been able for the first time to perform a 
complete and realistic analysis of the $\Pg\Pg\to 4\Pleptons$ process at high 
energies. Up to now, this channel has been considered only for Higgs boson 
discovery. Its behaviour has been thus analysed for kinematical configurations 
appropriate for the Higgs search, and at energy scales around the expected 
Higgs resonance. The new fit to the electroweak precision data has recently 
lowered the 95$\%$C.L. bound on the Higgs mass down to about 144 $\GeV$. If 
one takes this result as a reasonable indication, the scanned energy domain is 
quite narrow.

The purpose of this Letter is to extend the analysis at the $\TeV$ scale,
and show that the gluon-fusion channel constitutes a powerful probe into the 
EWSB physics, independently on the Higgs existence and discovery. Following a 
model-independent approach, we focus on the interaction between the produced 
weak-bosons. This interaction can be modified 
by the presence of new EWSB physics, which might appear at energy scales 
probed at the LHC or even larger. Having as a target the study of the 
sensitivity of the considered channel to possible new physics, we parametrize 
such a scenario by choosing the minimal realization, \ie the Standard Model 
with no Higgs. And, we compare the outcoming results with the predictions of
the SM with a light Higgs.

Our aim is to perform a complete and realistic analysis of the weak-boson 
pair production at the LHC, via the $\Pp\Pp\to 4\Pleptons$ process (see e.g.
Ref.~\cite{Haywood} for a review on its present status). 
In addition to the gluon-induced signal, we have to consider the background
coming from quark-induced contributions to the same final state, 
$\Pq\bar\Pq\to 4\Pleptons$. This background is overwhelming, but it can be 
heavily suppressed as shown later. 
     
Before starting our analysis, let us summarize the numerical setup. In 
computing partonic cross-sections, for the SM free parameters we use the input 
values \cite{Hagiwara:pw-mtop}:
\beq\label{eq:SMpar}
\begin{array}[b]{lcllcllcl}
\MW & = & 80.403\GeV, &
\MZ & = & 91.1876\GeV, \\
\Mt & = & 174.2 \GeV, &
\Mb & = & 4.4 \GeV,  \\
G_\mu &= & 1.16639 \times 10^{-5} \GeV^{-2}. & \\ 
\end{array}
\eeq
The weak mixing angle is fixed by $\sw^2=1-\MW^2/\MZ^2$. We adopt the 
$G_{\mu}$-scheme, which effectively includes higher-order contributions 
associated with the running of the electromagnetic coupling and the leading 
universal two-loop $\Mt$-dependent corrections. To this end we parametrize the 
LO matrix element in terms of the effective coupling 
$\alpha_{G_{\mu}}=\sqrt{2}G_{\mu}\MW^2\sw^2/\pi$. We moreover use the 
fixed-width scheme with $\GZ = 2.44506 \GeV$ and $\GW = 2.04685 \GeV$. As to 
parton distributions (PDF), we have chosen CTEQ6M \cite{cteq} at the 
factorization scale $Q=\MW$. We consider purely leptonic final states:
\begin{equation}\label{eq:states}
\Pp\Pp\to\nu_\Pl\Pl^+{\Pl^\prime}^-\bar\nu_{\Pl^\prime}~~~~~~~~
\Pl,\Pl^\prime=\Pe,\mu. 
\end{equation}
The signature is thus characterized by two isolated charged leptons plus 
missing energy. This channel includes the $\PW\PW$ production as intermediate 
state. In the parton model, the corresponding cross sections are described by 
the following convolution 
\beqar\label{eq:convolWW}
&&\rd\si^{\Pp\Pp}(P_1,P_2,p_f) =  
\int_0^1\rd x_1 \rd x_2 \sum_{q=\Pg,\Pu,\Pd,\Pc,\Ps}
\Bigl[\Phi_{\bar\Pq,\Pp}(x_1,Q^2)
\nl&&{}
\times\Phi_{\Pq,\Pp}(x_2,Q^2)\,\rd\hat\si^{\bar\Pq\Pq}(x_1P_1,x_2P_2,p_f)
+\Phi_{\bar\Pq,\Pp}(x_2,Q^2)
\nl&&{}
\times\Phi_{\Pq,\Pp}(x_1,Q^2)\times\rd\hat\si^{\bar\Pq\Pq}(x_2P_2,x_1P_1,p_f)
\Bigr]
\eeqar
where $p_f$ summarizes the final-state momenta, $\Phi_{i,h_i}$ is the PDF of 
parton $i$ in the incoming proton $h_i$ with momenta $P_i$, and $\hat\si^{ij}$ 
represent the partonic, colour and spin avaraged, cross sections. The 
$\hat\si^{ij}$ are calculated at LO, using the matrix elements for the 
complete processes
\beq\label{eq:SMpar}
\begin{array}[b]{lc}
g(p_1) + g(p_2)\to f_3(p_3) + f_4(p_4) + f_5(p_5) + f_6(p_6) & \\
\bar q_1(p_1) + q_2(p_2)\to f_3(p_3) + f_4(p_4) + f_5(p_5) + f_6(p_6) & \\
\end{array}
\eeq
where the arguments label the momenta $p_i$ of the external gluons and 
fermions. This means that 
we include the full set of Feynman diagrams, in this way accounting for the 
resonant di-boson production as well as the irreducible background coming from 
non-doubly resonant contributions.
Complete four-fermion phase spaces and exact kinematics are employed in our
calculation. 
\begin{table}\centering
\begin{tabular}{|c|c|c|}
\hline 
$\Mll^{\cut}$ & $\si_{\PBox}(g_{1,2,3})/\si_{\PBox}(g_{1,2})$ & 
$\si_{H,\PBox(g_3)}/[\si_H+\si_{\PBox(g_3)}]$\\
\hline
0 $\GeV$   & 1.16 & 0.57 \\ 
500 $\GeV$ & 5.35 & 0.09 \\ 
\hline
\end{tabular}
\caption {Relative size of individual contributions to the process 
$\Pp\Pp\to\nu_\Pe\Pe^+\mu^-\bar\nu_\mu$ as a function 
of the cut on the invariant mass of the charged lepton pair. 
Table entries are explained in the text. Standard cuts are applied.}
\vspace{-.3cm}
\label{ta:contributions}
\end{table}

For the experimental identification of the final state particles, we have 
implemented a general set of cuts appropriate for the LHC, and defined 
as follows:
\begin{itemize}
\item {lepton transverse momentum $\PT(\Pl^\pm )>20\GeV$},

\item {missing transverse momentum $\PTmiss> 25\GeV$}, 
  
\item {charged lepton rapidity $|y_\Pl |< 2$}, where
  $y_\Pl=-\log\left (\tan(\theta_\Pl/2)\right )$, and $\theta_\Pl$ is the
  polar angle of particle $\Pl$ (massless) with respect to the beam.
\end{itemize}
These are standard cuts, dedicated ones will be described at due time.

We begin our analysis by comparing the properties of the gluon-induced process 
in the low and high energy regime. An important difference is in the behavior 
of the top-bottom massive quark loop. At low energy, the contribution of the 
third quark generation to the box diagram is neglegible compared to the 
contribution of the first two generations. This behaviour changes drastically 
at high energies. The amplitude of the massive quark box grows with 
increasing energy, and becomes dominant. Moreover, it interferes strongly and 
destructively with the Higgs diagram. This is shown in 
Tab.~\ref{ta:contributions} as a function of the cut on the invariant mass
of the two charged leptons, $\Mll^{\cut}$. The second column presents the 
ratio between the full box cross section and the contribution of the first two 
generations. Whereas at low energies the top-bottom quark loop 
constitutes only the 16$\%$ of the total box cross-section, already for 
$\Mll > 500 \GeV$ (which means $E_{\cm}\simeq 1 \TeV$) it gets dominant by a 
factor 5 over the light quark generations. 
The third column shows instead the interference between the box 
graph mediated by the third generation quarks and the Higgs diagram.
With increasing energy, the interference gets heavily negative and gives rise
to a cancellation between the two amplitudes of about a factor 10.
 
An analogous feature is displayed by the loop-induced $\PZ$-boson pair 
production as discussed in Ref.\cite{Glover:1989}. This peculiar behaviour 
finds an explanation in the analitical expression of the matrix element. Both 
the Higgs and the massive quark box amplitudes squared exhibit indeed a 
logarithmic dependence on the
center-of-mass energy of the $\Pg\Pg\to 4\Pf$ process. Such an energy 
dependence can be interpreted as a relic of the much stronger energy 
dependence of the individual contributions to the on-shell process 
$\Pt\bar\Pt\to\PW^+\PW^-\to 4\Pf$, obtained by cutting on the internal 
lines of the loops appearing in the graphs of Fig.~\ref{fig:LOdiag} and 
related ones. The amplitude squared of the on-shell top-induced $\PW\PW$-pair 
production grows like $s=E_{\Pcm}^2$ in absence of the Higgs. 
The same energy dependence is shared by the additional Higgs contribution.
Gauge and Higgs amplitudes interfere destructively in order to preserve the 
perturbative unitarity of the theory. This feature is not washed out by the 
convolution of the top-induced subprocess with the gluon-induced quark loop. 
A di-logarithmic energy dependence of the individual graphs indeed survives, 
as mentioned above.   

\begin{table}\centering
\begin{tabular}{|c|c|c|c|}
\hline 
Setup & $\si(\Pq\bar\Pq )~(\fba )$ & $\si (\Pg\Pg)~(\fba )$ & $\si(\Pq\bar\Pq )/\si (\Pg\Pg)$ \\ 
\hline
no cuts        & 555.4 & 58.6 & 9.5 \\ 
Standard  cuts & 128.2 & 24.1 & 5.3 \\
Dedicated cuts & 5.2   & 3.0  & 1.7 \\
\hline
\end{tabular}
\caption {From left to right, SM $\Pq\bar\Pq$-background, 
$\Pg\Pg$-signal in the no-Higgs scenario, and their ratio for the process 
$\Pp\Pp\to\nu_\Pe\Pe^+\mu^-\bar\nu_\mu$ and different sets of cuts, as 
described in the text.}  
\vspace{-.3cm}
\label{ta:qqback}
\end{table}

\begin{figure}
  \unitlength 1cm
  \begin{center}
  \begin{picture}(16.,15.)
  \put(-2.5,-1){\epsfig{file=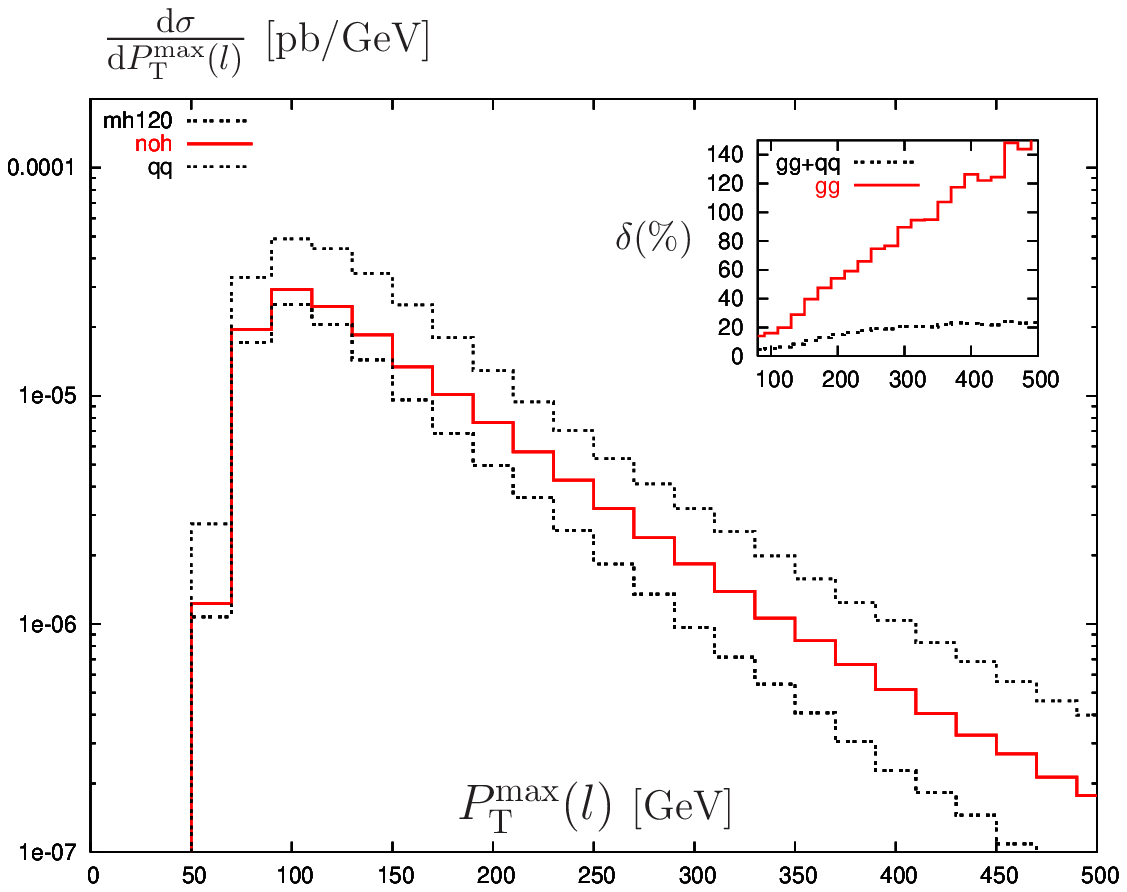,width=14cm}}
  \put(-2.5,-7){\epsfig{file=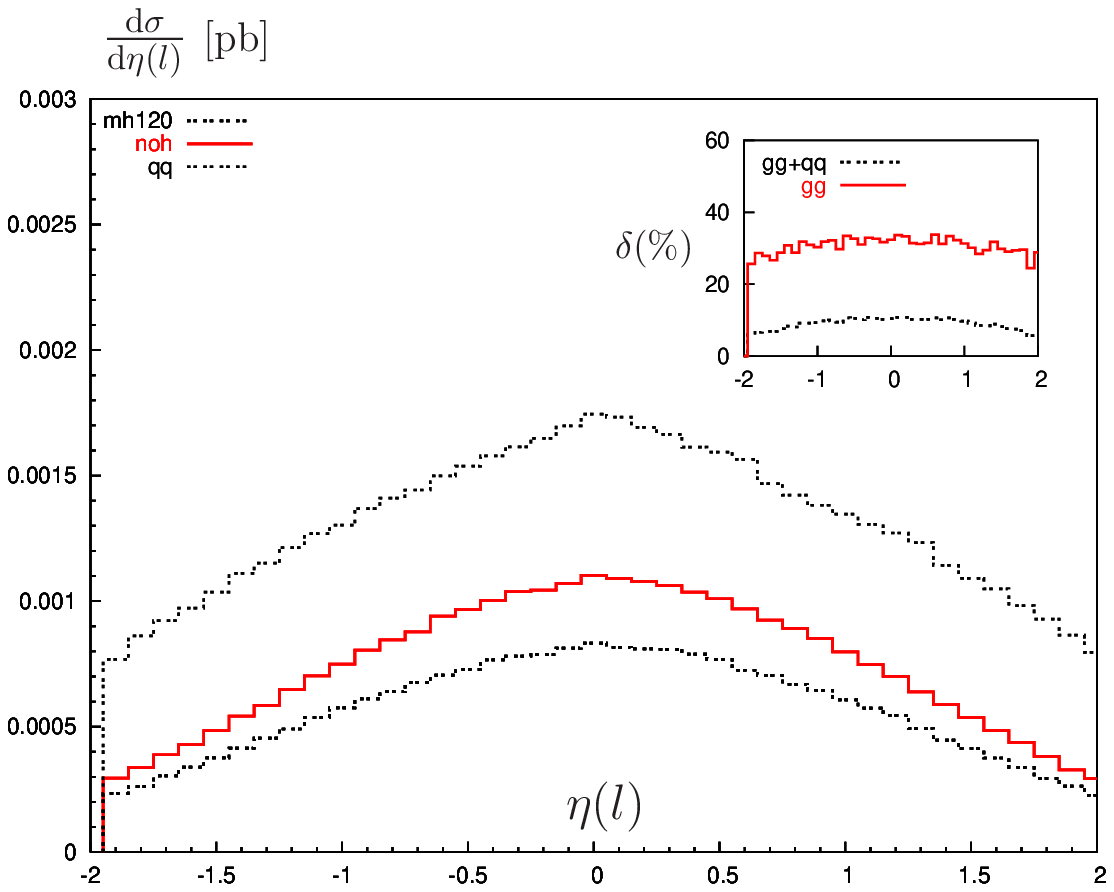,width=14cm}}
  \end{picture}
  \end{center}
\vspace{-3.7cm}
\caption{a) Distribution in the maximal transverse momentum of the charged 
leptons. b) Distribution in the rapidity of the negatively charged lepton. 
We consider the process $\Pp\Pp\to\nu_\Pe\Pe^+\mu^-\bar\nu_\mu$. From top to 
bottom, the three curves represent SM $\Pq\bar\Pq$-background, 
$\Pg\Pg$-signal with no Higgs, and $\Pg\Pg$-process in the SM ($M_H=120\GeV$). 
The inset plot gives the difference 
in percent between SM and noH scenarios for the gluon-induced (upper curve) 
and the full process (lower curve). Dedicated cuts are applied.}
\vspace{-.3cm}
\label{fig:distributions}
\end{figure}
We exploit this behaviour in order to quantify the sensitivity of the 
gluon-induced channel to possible new EWSB physics. We consider two benchmark 
scenarios: the SM with a light Higgs ($M_H=120 \GeV$) and the SM with no 
Higgs (noH). For a realistic assessment of the 
potential of the considered process, one has to take into account the full
background coming from the process $\Pq\bar\Pq\to 4\Pleptons$ and 
giving 
rise to the same final state. At first glance, this large contribution seems 
to bury away any possible new-physics signal. In absence of any cut, its cross 
section is a factor 10 bigger than the gluon-induced one. As shown in 
Tab.~\ref{ta:qqback}, standard cuts help in reducing it down to a factor 5.
But, still the discovery potential of the gluon-induced channel is largely 
spoiled. The only way to keep under control the large 
$\Pq\bar\Pq$-background is to exploit the pronounced differences shown 
in many variable distributions by this process compared to the loop-induced 
signal.

The most important kinematical difference is in the rapidity distribution of 
the final state leptons. The $\Pq\bar\Pq$-background tends to be 
produced at a larger rapidity, because of the harder distribution of the 
valence quarks. The second most important one lies in the spin state of the 
intermediate $\PW$-pair system. The signal we are interested in can be traced 
back to the production of longitudinal $\PW$-bosons, for it is this 
rate to be enhanced by possible new EWSB physics. Such a signal is expected 
to increase for high CM energies and large scattering angles of the produced 
$\PW$'s. In order to recover the lost sensitivity, we thus impose the following
kinematical constraints:
\begin{itemize}
\item {missing transverse momentum $\PTmiss> 80\GeV$},
\item {lepton azimutal angle difference 
$\Delta\phi (\Pl^+\Pl^{\prime -})>60^o$},
\item {charged lepton rapidity difference $\Delta y(\Pl\Pl^\prime )< 2$}, 
where $\Delta y(\Pl\Pl^\prime )=|y_\Pl-y_{\Pl^\prime}|$
\item {charged lepton opening-angle $\Pcos\theta (\Pl\Pl^\prime )>-0.98$.}
\end{itemize}
With this choice, we refer to as dedicated cuts, the 
$\Pq\bar\Pq$-background becomes of the same order of magnitude as the 
$\Pg\Pg$-signal, as shown in the last row of Tab.~\ref{ta:qqback}, partially 
recovering the lost sensitivity. The imposed cuts also select large energies 
($E_{\cm}\geq 300\GeV$) and angles, as of interest.  
\begin{figure}
  \unitlength 1cm
  \begin{center}
  \begin{picture}(16.,15.)
  \put(-2.5,-1){\epsfig{file=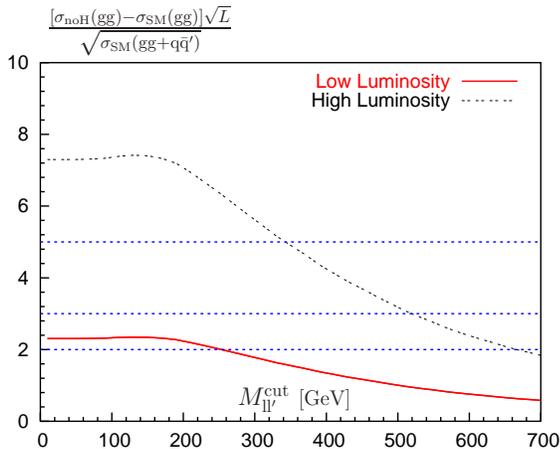,width=14cm}}
  \end{picture}
  \end{center}
\vspace{-9.7cm}
\caption{Signal over background ratio, for the process 
$\Pp\Pp\to\nu_\Pl\Pl^+\Pl^{\prime -}\bar\nu_{\Pl^\prime}$. We sum over $\Pe$ 
and $\mu$, and consider two luminosity values: $L=10\fba^{-1}$ (lower curve) 
and $L=100\fba^{-1}$ (upper curve). Dedicated cuts are applied.}
\label{fig:stdev}
\end{figure}

With these results at hand, we are ready to present the first estimate of the 
full $\Pp\Pp\to 4\Pleptons$ process at high energy scales. We call it 
estimate, because we do not include NLO QCD corrections available for the 
$\Pq\bar\Pq$-background (see for instance Ref.~\cite{NLOQCD}), but 
still missing for the gluon-induced signal. We show results for the LHC at 
$E_{\cm}=14\TeV$. In Fig.~\ref{fig:distributions}, we plot two sample 
distributions in energy and angle, comparing the gluon-induced 
signal in the no-Higgs scenario ($\noH$) with the SM prediction given by the 
$\Pq\bar\Pq$-background plus the $\Pg\Pg$-process with a light Higgs 
($M_H=120\GeV$). The inset plots show that the difference between the two 
benchmark scenarios is enhanced at high energies and large angles of the 
outgoing charged leptons. Such a difference would be extremely pronounced 
if we considered only the gluon-fusion contribution, 
$\delta = \si_{\noH}(\Pg\Pg)/\si_{\SM}(\Pg\Pg)-1$ (upper curves). The 
$\Pq\bar\Pq$-background reduces it, still preserving a difference, 
$\delta = \si_{\noH}(\Pg\Pg)/\si_{\SM}(\Pg\Pg+\Pq\bar\Pq )-1$, up to 
20$\%$ (lower curves).     
\begin{table}\centering
\begin{tabular}{|c|c|c|c|}
\hline 
$\Mll^{\cut} (\GeV)$ & $N_{\noH}(\Pg\Pg)$ & $N_{\SM}(\Pg\Pg)$ & $N_{\SM}(\Pq\bar\Pq )$ \\ 
\hline
0   & 751 & 572 & 1323 \\ 
250 & 147 & 77  & 307  \\
500 & 25  & 9   & 64   \\
\hline
\end{tabular}
\caption {We consider the process $\Pp\Pp\to\nu_\Pe\Pe^+\mu^-\bar\nu_\mu$. 
From left to right, number of events for $\Pg\Pg$-process with no 
Higgs, SM $\Pg\Pg$-process ($M_H=120\GeV$), and SM 
$\Pq\bar\Pq$-background for $L=100\fba^{-1}$. Dedicated cuts are applied.}  
\vspace{-.3cm}
\label{ta:events}
\end{table}

In order to assess the sensitivity of the considered channel to possible 
new physics, one has to estimate the statistical significance of such effects.
We naively derive it from their comparison with the statistical error expected 
at the LHC. In Fig.~\ref{fig:stdev}, we plot the signal over background ratio
as a function of $\Mll^{cut}$. We 
consider the two envisaged values of the luminosity, $L=10\fba^{-1}$ and 
$L=100\fba^{-1}$, corresponding to the low- and high-luminosity run. The three 
horizontal lines are the 2, 3 and 5 standard deviation reference values. 
Fig.~\ref{fig:stdev} shows that at high luminosity, one can 
reach 2$\sigma$-effects and more over almost the entire energy range. Even in 
the low-luminosity run, one could explore sensitivity up to scales of the 
order of 500 $\GeV$.      
The number of estimated events at high-luminosity is given in 
Tab.~\ref{ta:events} as a function of $\Mll^{cut}$.

To conclude, we have provided the first complete study of the 
$\Pp\Pp\to\nu_\Pl\Pl^+\Pl^{\prime -}\bar\nu_{\Pl^\prime}$ process at 
${\cal O}(\alpha_s^2\alpha_{em}^4)$ in the high energy domain. This channel 
is found to have strong potential for probing the nature of EWSB physics at 
the LHC. A final statement should include detector response and systematics; 
this goes beyond our purpose.

T. Binoth and M. Ciccolini are gratefully acknowledged for valuable 
discussions and their help in using the $\tt gg2WW$ code. This work was 
supported by MIUR under contract Decreto MIUR 26-01-2001 N.13 and contract 
2006020509$_0$04, and by the European Community's MRTN under contract 
MRTN-CT-2006-035505.


\end{document}